\documentclass{ewic}

\usepackage[utf8]{inputenc}
\usepackage[english]{babel}
\usepackage[cmex10]{amsmath}
\usepackage{amssymb}
\usepackage{mathpartir}
\usepackage{xcolor}
\usepackage{multirow}
\usepackage{alltt}
\usepackage{url}
\usepackage{xspace}
\usepackage{enumerate}
\usepackage{bbm}
\usepackage{stmaryrd}
\usepackage{url}
\usepackage{tabularx}
\usepackage{listings}
\usepackage[font=footnotesize,caption=false]{subfig}
\usepackage{floatflt}
\usepackage{tikz}
\usepackage{longtable}
\usepackage{hyperref}
\usepackage{appendix}
\usepackage{tabularx}
\usepackage{graphicx}

\usetikzlibrary{petri}
\usetikzlibrary{positioning}
\usetikzlibrary{arrows}
\usepgflibrary{shapes}
\usetikzlibrary{decorations.pathreplacing}
\usetikzlibrary{calc}
\newif\iflong
\newcommand{\code}[1]{\ensuremath{{\textsf{\textbf{#1}}}}}%

\newcommand{\ilow}{\mathop{\downarrow}\kern -0.4ex}
\newcommand{\iup}{\mathop{\uparrow}\kern -0.4ex}
\newcommand{\eqdef}{\smash[t]{{\overset{\text{def}}{=}}}}

\def\Events{T}
\def\event{t}
\let\s\sigma

\newcommand{\real}{\mathbb{R}}
\newcommand{\realp}{\real^{+}}

\newcommand{\nat}{\mathbb{N}}
\def\until{\mathrel{\mathbf{{U}}}}
\let\ltlexist\Diamond
\let\ltlall\oblong

\def\such{\ .\ }

\def\obs{O}

\let\imply\Rightarrow

\def\dom{\mathop{\mathsf{dom}}}

\let\comp\parallel 

\def\P{P}
\def\T{T}

\def\B{B}
\def\F{F}

\def\sync{\T_{sync}}
\def\imm{\T_{imm}}
\def\lsync{L_{sync}}

\def\set#1{\,{\{#1\}}}

\tikzstyle{transition} =[very thick, rectangle, draw, inner xsep=2mm, inner ysep=0.75mm]
\tikzstyle{vtransition}=[very thick, rectangle, draw, inner ysep=2mm, inner xsep=0.75mm]
\tikzstyle{place}=[circle, draw, minimum size=3ex]
\tikzstyle{every label}=[font=\sf\footnotesize]
\def\lbl#1{\textsf{#1}} 
\tikzstyle{pre}+=[>=stealth]
\tikzstyle{post}+=[>=stealth]
\tikzstyle{readarc}=[pre, >=*, shorten <=0pt]
\tikzstyle{prio}=[draw, ->, orange, >=stealth, shorten >=1.25pt, shorten <=1.25pt, densely dashed]
\tikzstyle{poids}=[font=\scriptsize\sf]
\tikzstyle{action}=[rectangle, draw, color=black!80, thin, dotted, inner xsep=0.4ex, inner ysep=0.1ex]
\newcommand{\postact}[2][]{
{\begin{tikzpicture}[baseline, #1]
\node[action, anchor=text, #1] at (0,0) {\code{\scriptsize\strut\code{act:} {#2}}} ;
\end{tikzpicture}}}
\newcommand{\preact}[2][]{
{\begin{tikzpicture}[baseline, #1]
\node[action, anchor=text, #1] at (0,0) {\code{\scriptsize\strut\code{pre:} {#2}}} ;
\end{tikzpicture}}}

\tikzstyle{gil search}=[dashed, ->, >=triangle 60]
\tikzstyle{gil strong search}=[gil search, ->>]
\tikzstyle{gil context}=[{[-)}]
\tikzstyle{gil close context}=[{[-)}]
\tikzstyle{gil open context}=[{]-)}]
\tikzstyle{gil dotted context}=[loosely dotted,-]
\tikzstyle{gil open search}=[gil search, {]->}]
\tikzstyle{gil close search}=[gil search, {[->}]
\tikzstyle{gil time context}=[|-|]
\tikzstyle{gil open strong search}=[gil search, {]->>}]
\tikzstyle{gil close strong search}=[gil search, {[->>}]
\tikzstyle{gil segment}=[|-|]
\tikzstyle{gil exists}=[draw,kite,midway,kite vertex angles=60, inner sep=0.4ex]
\tikzstyle{gil always}=[gil exists, rectangle, inner ysep=0.8ex]
\tikzstyle{gil boite}=[dotted, thick, rounded corners=4pt]
\tikzstyle{gil interval}=[decorate,decoration={brace}]
\tikzstyle{every pin}=[pin distance=0.4ex]
\tikzstyle{every pin edge}=[draw,-]
\tikzstyle{audessus}=[anchor=south east, inner sep=0em, yshift=1.2ex]
\tikzstyle{audessous}=[anchor=north east, inner sep=0em, yshift=-1.2ex]
\tikzstyle{timint}=[midway,above=-0.5ex]
\tikzstyle{snippet}=[rectangle, rounded corners=0pt, inner ysep=0.15ex, inner xsep=1ex, semithick,
                     densely dotted, draw=black, text width=]

\newcommand{\pop}[1]{\ensuremath{\mathop{\text{\textit{{#1}\,}}}}}
\newtheorem{theorem}{Theorem}
\newtheorem{definition}{Definition}

\newtheorem{lemma}{Lemma}

\tikzstyle{patterncommon}=[anchor=text, draw=black]

\tikzstyle{patterntitle}=[patterncommon,rectangle,rounded corners=2pt, inner ysep=0.75ex, inner xsep=3ex]
\tikzstyle{patternexample}=[patterncommon,rectangle,draw=none, inner ysep=0.25ex, inner xsep=0.8ex]

\newcommand{\patterntitle}[1]{
\tikz[baseline]{\node[patterntitle] {\textbf{\textsf{#1}}} ; }
}

\newcommand{\fichepattern}[2]{
\vbox{
 \noindent\leftline{\patterntitle{#1}}\\[0.2em]
 \noindent\textit{#2}\medskip
}
}

\begin{document}

\runningheads{Abid $\bullet$ Dal Zilio $\bullet$ Le Botlan}{A Verified
  Approach for Checking Real-Time Specification Patterns}

\conference{Proceedings of \dots}

\title{A Verified Approach for Checking Real-Time Specification
  Patterns}

\authorone{Nouha Abid \qquad Silvano Dal Zilio \qquad  Didier Le Botlan\\
CNRS, LAAS, 7 avenue du colonel Roche, F-31400 Toulouse\\
Univ de Toulouse, INSA, LAAS, F-31400 Toulouse, France\\
\email{\{nabid, dalzilio, dlebotla\}@laas.fr}}

\begin{abstract}


  We propose a verified approach to the formal verification of timed
  properties using model-checking techniques. We focus on properties
  expressed using real-time specification patterns, which can be
  viewed as a subset of timed temporal logics that includes properties
  commonly found during the analysis of reactive systems.  Our
  model-checking approach is based on the use of observers in order to
  transform the verification of timed patterns into the verification
  of simpler LTL formulas. While the use of observers for
  model-checking is quite common, our contribution is original in
  several ways. First, we define a formal framework to verify that our
  observers are correct and non-intrusive. Second, we define different
  classes of observers for each pattern and use a pragmatic approach
  in order to select the most efficient candidate in practice. This
  approach is implemented in an integrated verification tool chain for
  the Fiacre language.
\end{abstract}


\maketitle
\pagestyle{empty}
\thispagestyle{empty}
\section{Introduction}
 distinctive feature of real-time systems is to be subject to severe
time constraints that arise from critical interactions between the
system and its environment. Since reasoning about real-time systems is
difficult, it is important to be able to apply formal validation
techniques early during the development process and to define formally
the requirements that need to be checked.

In this work, we follow a classical approach to model checking: (1) we
use a high-level language to describe a model of the system; (2) we
use a logical-based formalism to express requirements on the system;
and (3) the verification consists in compiling the system's model and
requirements into a low-level model for which we have the appropriate
theory and the convenient tooling. We propose a new treatment for this
traditional approach. In particular, for point (2), we focus on a
dense real-time model and we use \emph{real-time patterns} for the
specification of the system instead of timed extensions of temporal
logic. Our patterns can be interpreted as a real-time extension to the
specification patterns of~\citet{ppsfsv1999}. Time patterns can be
used to express constraints on the timing as well as the order of
events, such as the compliance to deadline or minimum time bounds on
the delay between events.  Concerning verification, point (3), we work
with Time Transition Systems (see Sect.~\ref{sec:time-trans-syst}), an
extension of Time Petri Nets with data variables and priorities.




Our first contribution is to propose a decidable verification method
for checking real-time patterns on Time Transition Systems (TTS). The
method is based on the use of observers and model-checking techniques
in order to transform the verification of patterns into the
verification of simpler LTL formula. 
Our observers are proved correct and non-intrusive, meaning that
they compute the correct answer and have no impact on the system
under observation. This is why we say our approach is verified. The
formal framework we have defined is not only useful for proving the validity
of formal results but also to check the soundness of optimisation
in the implementation.

\begin{figure*}
\centering
\begin{tabular}[c]{l}
\includegraphics[height=6cm]{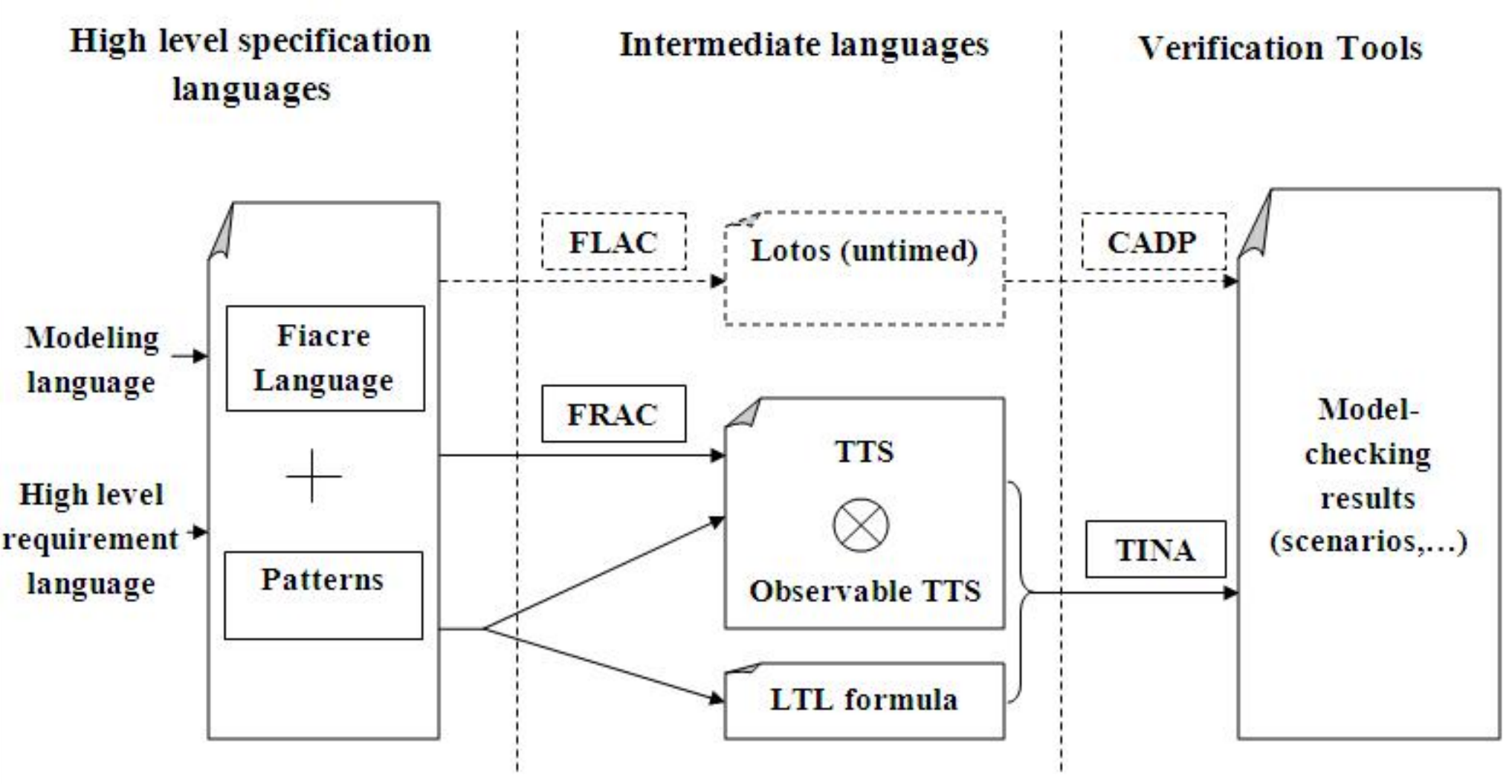}
\end{tabular}
\caption{The global verification tool chain}
\label{fig:intro}
\end{figure*}

Our second contribution is to provide a reference implementation for
these timed patterns. The complete framework defined in this paper has
been integrated into a verification tool chain for
Fiacre~\citep{filfmvte2008}, a high-level modelling language that can
be compiled to TTS. Fiacre can be used as input language for two
verification toolboxes: TINA, the TIme Petri Net Analyzer
tool set~\citep{tina}, and CADP~\citep{cadp}. In our tool chain
(described in Fig.~\ref{fig:intro}) a Fiacre specification is combined
with patterns and compiled into a TTS model using the Frac compiler (the Fiacre 
language compiler).
Then the model can be checked using the TINA toolbox. This is not a
toy example. Indeed, Fiacre is the intermediate language used for
model verification in Topcased~\citep{ttptosfcasd2006}, an Eclipse
based toolkit for critical systems, where it is used as the target of
model transformation engines from various languages, such as SDL, BPEL
or AADL~\citep{aadl2fcr}. Therefore, through the connection with
Fiacre, we can check timed patterns on many different modelling
languages.


Due to space limitations, we only give a partial descriptions of our
timed patterns and give only part of our theoretical results. A
complete catalogue of timed specification patterns is given
in~\cite{FRP11}, while the complete formal framework is defined in a
long version of this paper~\citep{VRTS11}.

For the purpose of this work, we focus on a simple \emph{deadline
  pattern}, named \pop{leadsto}, and define different classes of
observers that can be used to check this pattern. We define observers
for the \pop{leadsto} patterns that are based on the monitoring of
places or transitions. In addition to these two traditional kind of
observers, we propose a class of TTS observers that monitor data
modifications. The goal is to choose the most efficient observer in
practice.  We give some experimental results on the impact of the
choice of observer on the size of the state graphs that need to be
generated---that is on the space complexity of our verification
method---and on the verification time. The goal of this particular
study is not to define a method for automatically generating an
observer from a pattern. Instead, we define a set of possible
observers that are compared in order to choose the best one in
practice.

\subsubsection*{Outline} 
The paper is organised as follows. We start by introducing our formal
framework in Sect.~\ref{sec:time-trans-syst}. This section is useful
to define the notion of composition and non-interference for our
observers. In Sect.~\ref{sec:prop-extens-}
and~\ref{sec:real-time-properties}, we describe a subset of our
real-time specification patterns and the verification framework. We
describe the implementation of our tool chain and give some
experimental results on the use of the \pop{leadsto} pattern in
Sect.~\ref{sec:experimental-results}. We conclude with a review of the
related work, an outline of our contributions and some perspectives on
future work.

\section{Formal framework}
\label{sec:time-trans-syst}

We define some formal notations that are used in the remainder of this
paper. In our approach, the observers and the systems are presented as
Time Transition System (TTS), an extension of Time Petri Nets (TPN)
\citetext{see e.g.~\citealp{merlin}} with data variables and
priorities. Our formal framework is based on the work
of~\citet{OCTPN}, where the authors define formally the composition of
two TPN. Their presentation has been extended to the full TTS model
in~\citet{VRTS11}.

The notion of composition is important in our work since we use TTS
models for both the system and the observer and, for verification, we
use TTS composition to graft the system with the observer.

This section is organised as follows: first, we introduce informally 
a TTS example. Then, we give a formal definition
of TPN following the presentation of~\citet{OCTPN}, which is then extended
to TTS. The semantics of TTS is defined using sets of
timed traces. Finally, we define the composition of two TTS. 

\subsection{Informal Presentation of the TTS Model}
\label{sub:introduction}
\begin{figure}
\centerline{
\begin{tikzpicture}[node distance=4.5ex and 7.5ex, label distance=-0.5ex]
\node[place, label=below:$s_0$, tokens=1] (s0) {} ;
\node[vtransition, right=of s0, label=above:click] (t1) {}
  edge[pre] (s0) ;
\node[place, label=below:$s_1$, right=of t1]  (s1) {}
  edge[pre] (t1) ;
\node[vtransition, right=of s1, label=below:{$[1;1]$}, label=above:{$\tau$}] (t2) {}
  edge[pre] (s1) ;
\node[place, right=of t2, label=below:$s_2$] (s2) {}
  edge[pre] (t2) ;
\node[vtransition, below=6ex of s1, label=above:double,
label=below:\begin{tabular}{c}\preact{dbl == true}\\\postact{dbl := false}\end{tabular}] (t3) {}
  edge[pre, in=240, out=0] (s2) 
  edge[post, in=-60, out=180] (s0) ;
\node[transition, above=3.5ex of s1, label=left:click, label=above:\postact{dbl := true}] (t4) {}
  edge[readarc] (s1) ;
\node[transition, above=7.5ex of t4, label=below:single,
label=above:\begin{tabular}{c}\postact{dbl := false}\\\preact{dbl == false}\end{tabular}] (t5) {}
  edge[pre, in=120, out=0] (s2) 
  edge[post, in=60, out=180] (s0) ;
\draw[prio] (t2) -- (t4) ;
\end{tikzpicture}}
\caption{The double-click example in TTS}
\label{fig/dble-TTS}
\end{figure}
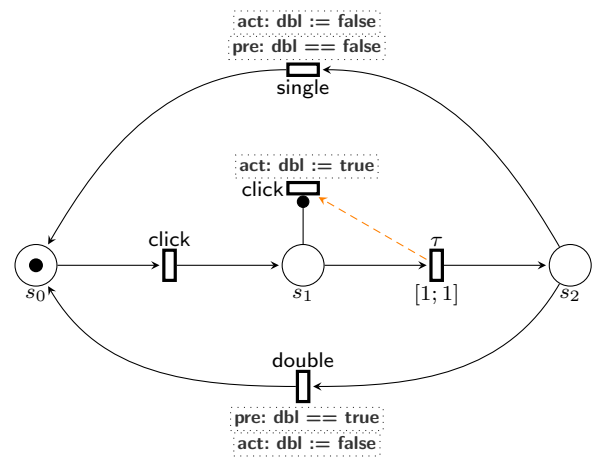

We introduce next a graphical syntax of TTS using a simple
example that models the behaviour of a mouse button with
double-clicking, as pictured in Fig.~\ref{fig/dble-TTS}. 
The behaviour, in this case, is to emit the event
\lbl{double} if there are more than two \lbl{click} events in
\emph{strictly less} than one unit of time (u.t.).

Ignoring at first side conditions and side effects (the \pop{pre} and
\pop{act} expressions inside dotted rectangles), the TTS in
Fig.~\ref{fig/dble-TTS} can be viewed as a TPN with one token in place
$s_0$ as its initial marking. From this ``state'', a \lbl{click}
transition may occur and move the token from $s_0$ to $s_1$. With this
marking, the internal transition $\tau$ is enabled and will fire after
exactly one unit of time, since the token in $s_1$ is not consumed by any other transition.
Meanwhile, the transition labeled \lbl{click} may fire one or more
times without removing the token from $s_1$, as indicated by the
\emph{read arc} (arcs ending with a black dot). After exactly one unit
of time, because of the  priority arc (a
dashed arrow between transitions),
the \lbl{click} transition is disabled until
the token moves from $s_1$ to $s_2$.

Data is managed within the \pop{act} and \pop{pre} expressions that
may be associated to each transition.  These expressions may refer to
a fixed set of variables that form the \emph{store} of the TTS. Assume
$t$ is a transition with guards \pop{act}$_t$ and \pop{pre}$_t$. In
comparison with a TPN, a transition $t$ in a TTS is enabled if there
is both: (1) enough tokens in the places of its pre-condition; and (2)
the predicate \pop{pre}$_t$ is true. With respect to the firing of
$t$, the main difference is that we modify the store by executing the
action guard \pop{act}$_t$. For example, when the token reaches the
place $s_2$ in the TTS of Fig.~\ref{fig/dble-TTS}, we use the value of
the variable \lbl{dbl} to test whether we should signal a double click
or not.

\subsection{Labeled Time Petri Nets and Time Transition Systems}
 \label{sub:TPN-def}
 Labeled Time Petri Nets (or TPN) extend Time Petri
 Nets~\citep{merlin} with an action alphabet and a function labelling
 the transitions with those actions.

 Notation~: Let $I^+$ be the set of nonempty real intervals with non
 negative rational endpoints.  For $i \in I^+$ , the symbol
 $\ilow i$ denotes the left end-point of the interval $i$ and
 $\iup i$ its right end-point, if $i$ is bounded, or $\infty$
 otherwise. We use $\nat$ to denote the set of non negative integers.

 \begin{definition}\label{def:TPN} A labeled Time Petri Net (or TPN)
   is a 8-tuple $(\P, \T, \B, \F, M_{0}, I_{s}, \sum, L)$ in which:
 \begin{itemize}
 \item $\P$ is a finite set of places $p_{i}$;

 \item $\T$ is a finite set of transitions $t_{i}$;
   
 \item $\B$ is the backward incidence function\\
   \begin{tabular}{c}         
     $\B:\T \rightarrow \P \rightarrow \nat;$\\ 
   \end{tabular}
 
 \item $\F$ is the forward incidence function\\
   \begin{tabular}{c}    
     $\F:\T \rightarrow \P \rightarrow \nat;$
   \end{tabular}
   
 \item $M_{0}$ is the initial marking function \\
   \begin{tabular}{c}
     $M_{0}:\P \rightarrow \nat;$
   \end{tabular}
 \item $I_{s}$ is a function called the static interval function\\
   \begin{tabular}{c}
     $I_{s}:\T \rightarrow I^+;$
   \end{tabular}\\
   Function $I_{s}$ associates a temporal interval $I_{s}(t) \in I^+$ with 
   every transition of the system. $\ilow I_{s}(t)$ and $\iup I_{s}(t)$
   are called the static earliest and latest firing times of
   $t$, 
   respectively.
   Assuming that a transition t became enabled at time $\tau$,
   then 
   $t$ cannot fire before $(\tau + \ilow I_{s}(t))$ 
   and no later than  $(\tau + \iup I_{s}(t))$ unless
   disabled by firing some other transition.
   
 \item $\sum$ is a finite set of actions, or labels, not containing
   the silent action $\varepsilon$;

 \item $L : \T \rightarrow \sum \cup \{\varepsilon\}$ is a transition
   labelling function.
 \end{itemize}
 A marking is a function $M : \P \rightarrow \nat$ that records the
 current (dynamic) value of the places in the net, as transitions are
 fired. The transition $t \in \T$ is enabled by $M$ iff $(M \geqslant
 B(t))$.  The dynamic interval function $I : T \rightarrow I^+$ is a
 mapping from transitions to time intervals. The dynamic interval
 function is used to record the current timing constraints associated
 to each transition, as time passes.
\end{definition}

 
A transition $t$ can fire from $(M, I)$ if $t$ is enabled at $M$ and
instantly fireable, that is $0 \in I(t)$.  In the target state, the
transitions that remained enabled while $t$ is fired ($t$ excluded)
keep their time interval, the intervals of the others (newly enabled)
transitions are set to their respective static intervals. Together
with those ``discrete'' transitions, a time Petri Net adds the ability
to model the flowing of time. A transition of amount $d$
(i.e. taking $d$ time units) is possible iff $d$ is less than $\iup
I(t)$ for all enabled transitions $t$.

The definition of TTS is a natural extension of TPN that takes
variables and priorities into account.  Details are presented
in~\citet{VRTS11}.

\begin{definition}[Timed traces]
\label{def/timed-trace}
A timed trace $\s$ is a possibly infinite sequence of events 
$\event \in \Events$ and duration $d$ with $d \in \realp$.
Formally, $\s$ is a partial mapping from $\nat$ to $\dot\Events
= \Events \cup \set{d \mid d \in \realp}$ such that
$\s(i)$ is defined whenever $\s(j)$ is defined and $i \leq j$.

The domain of $\s$ is written $\dom\s$. If $\dom\s$ is finite, the
\emph{duration} of $\s$, denoted $\Delta(\s)$, is the sum of the
delays in $\s$, that is $\sum_{i \mid \s(i) \in \realp} \s(i)$.
\end{definition}

The semantics of a TPN (resp. TTS) is the set of its timed traces.
(see details in~\citet{VRTS11}).

\subsection{Composition of TTS and Timed Traces }
\label{sub:semantics}
We study the composition of two TTS and consider the relation between traces of
the composed system and traces of both components. This operation is
particularly significant in the context of this work, since both the system and the
observer are TTS and we use composition to graft the latter to the former.
In particular, we are interested in conditions ensuring that the
behaviour of the observer does not interfere with the behaviour of the
observed system.

The ``parallel composition'' of labeled Petri nets is a fundamental
operation that is used to model large systems by incrementally
combining smaller nets. Basically, the composition of two labeled TPN
$N_1$ and $N_2$ is a labeled net $N \eqdef (N_1 \comp N_2)$ such that:
the places of $N$ is the cartesian product of the places of $N_1$ and
$N_2$, and the transitions of $N$ is the fusion of the transitions in
$N_1$ and $N_2$ that have the same label. A formal definition for the
composition of two TPN is given in~\citet{OCTPN}.  Composition of TTS
is basically the same~\citep{VRTS11}, with the noticeable restriction
that transitions which have priority over other transitions may not be
synchronised across components. This is required to ensure the
compositionality theorem, which we introduce below.

In the same way, we can define the composition of timed traces as an
operation that builds a timed trace $\s_1 \comp \s_2$ from two traces
$\s_1$ and $\s_2$. The trace $\s_1 \comp \s_2$ is obtained by merging
the events with the same labels. This operation is well-defined for
pairs of \emph{composable traces}.  Let $N_1$ (resp. $N_2$) be a TPN,
and $\s_1$ (resp. $\s_2$) one of its traces.  We say that $\s_1$ and
$\s_2$ are \emph{composable} iff $\dom{\s_1} = \dom{\s_2}$, and for
all $i \in \dom{\s_1}$, (1) $\s_1(i) = d \wedge d \in \realp \imply
\s_2(i) = d$, and (2) $\s_1(i) = t \wedge t \in T \imply L(\s_1(i)) =
L(\s_2(i))$.

The compositionality theorem states that the behaviour of the composed
system (expressed as a set of timed traces) is a subset of the
behaviour of both components. In other terms, composing a system with
an observer cannot generate new behaviour.

\begin{theorem}[Compositionality]
\label{prop/composition}
Let $N_1$ and $N_2$ be two TTS and $N = N_1 \comp N_2$ be their
composition.  Then, for every timed trace $\s$ of $N$, there exist two
timed traces, $\s_1$ and $\s_2$, such that: (1) $\s_i$ is a trace of
$N_i$ for $i \in 1..2$ and (2) $\s = \s_1 \comp \s_2$.
\end{theorem}

In the compositionality theorem, the trace $\s_1$ (resp. $\s_2$) is
obtained from $\s$ by ``erasing'' all transitions of $N_2$
(resp. $N_1$).  Due to lack of space, we omit the proof here and
invite the reader to consult~\citet{VRTS11}.

\section{Real-Time Specification Patterns}
\label{sec:prop-extens-}

We have defined in~\citet{FRP11} a set of specification patterns
that can express constraints on the delays between the occurrences of
two events or on the duration of a given condition. In our context, the event of a model can be: a transition that
is fired; the system entering or leaving a state; a change in the
value of variables; \dots The advantage of
proposing predefined patterns is to provide a simple formalism to
non-experts for expressing properties that can be directly checked
with our verification tool chain. Our patterns can be viewed as a
real-time extension of Dwyer's~\citeyearpar{ppsfsv1999} specification
patterns. In his seminal work, Dwyer shows through a study of 500
specification examples that 80\% of the temporal requirements can be
covered by a small number of ``pattern formulas''. We follow a similar
philosophy and define a list of patterns that takes into account
timing constraints. At the syntactic level, this is mostly obtained by
extending Dwyer's patterns with two kind of \emph{timing modifiers}:
(1) $P$ \pop{within} $I$, which states that the delay between two
events declared in the pattern $P$ must fit in the time interval $I$;
and (2) $P$ \pop{lasting} $D$, which states that the condition
defined by $P$ must hold for at least duration $D$. For example, we
define a pattern {\pop{Present} $A$ \pop{after} $B$ \pop{within}
  $]0, 4]$} to express that the event $A$ must occur within 4 unit of
time of the first occurrence of event $B$, if any, and not
simultaneously with it. Although seemingly innocuous, the addition of
these two modifiers has a great impact on the semantics of patterns
and on the verification techniques that are involved.


We describe our patterns using a hierarchical classification borrowed
from~\citet{ppsfsv1999}, with patterns arranged in categories such as
universality, absence, response, etc. In the following, we give some
examples of \textit{absence} and \textit{response} patterns based on
the TTS example of Fig.~\ref{fig/dble-TTS}. Each of these patterns can
be checked using our tool chain. A complete catalogue of patterns, with
their formal definition, is given in~\citet{FRP11}. In this section,
we focus on the ``response pattern with delay'', to give an example of
how patterns can be formally defined and to explain our different
classes of observers.

\subsection{Absence Pattern with Delay} 

This category of patterns is used to specify delays within which
activities must not occur. A typical pattern in this category is:
\[\tag{absent} \pop{absent}
E_2 \pop{after} E_1 \pop{for interval} [d_1; d_2]~,
\]
which asserts that a transition (labeled with) $E_2$ cannot occur
between $d_1$ and $d_2$ units of time after the first occurrence of a
transition $E_1$. An example of use for this pattern would be the
requirement that we cannot have two double clicks in less than~$2$
units of time (u.t.), that is: \pop{absent} \code{double} \pop{after}
\code{double} \pop{for interval} $[0; 2]$. (This property is not true
for our example in Fig.~\ref{fig/dble-TTS}.) A more contrived example
is to require that if there are no single clicks in the first $10$
u.t. of an execution then there should be no double clicks at
all. This requirement can be expressed using the composition of two
absence patterns using the implication operator and the reserved
transition \code{init} (that identifies the start of the system):
\[
\begin{array}{l}
  \big(\pop{absent} \code{single} \pop{after} \code{init}
   \pop{for interval} [0;10]\big)\\
  \quad \Rightarrow \big(\pop{absent} \code{double}
  \pop{after} \code{init}  \pop{for interval} [0;\infty[ \big)~.
\end{array}
\]


\subsection{Response Pattern with Delay} 

This category of patterns is used to express that some (triggering)
event must always be followed by a given (response) event within a
fixed delay of time. The typical example of response pattern states
that every occurrence of a transition labeled with $E_1$ must be
followed by an occurrence of a transition labeled with $E_2$ within a
time interval $I$. (We consider the first occurrence of $E_2$ after
$E_1$.)
\[\tag{leadsto} E_1 \pop{leadsto} E_2 \pop{within} I~.\]
For example, using a disjunction between transition labels, we can
bound the time between a \code{click} and a mouse event with the
pattern: \code{click} \pop{leadsto} $(\code{single} \vee
\code{double})$ \pop{within} $[0,1]$.


\subsection{Other Examples of Patterns}

To give a feel of the expressiveness of our patterns, we briefly
describe some other examples. For each pattern, we give just a textual
definition. In each example, $E_1$, $E_2$ and $E_3$ refer to events in
the system and $d_1$ (resp. $d_2$) stand for the left end-point (resp.
right end-point) of the time interval $I$.\\ 

\fichepattern
{\code{Present} $E_1$ \code{after} $E_2$ \code{within} $I$}
{Predicate $E_1$ must hold
between $d_1$ and $d_2$ u.t after the first occurrence
of $E_2$. The pattern is also satisfied if $E_2$ never holds.
}

\fichepattern
{\code{Present first} $E_1$ \code{before} $E_2$ \code{within} $I$}
{The first occurrence of $E_1$ should be between $d_1$ and $d_2$
  u.t. before the first occurrence of~$E_2$. The pattern also holds if
  $E_2$ never occurs.}

\fichepattern
{\code{Present} $E_1$ \code{lasting} $D$}
{Starting from the first occurrence when the predicate $E_1$ holds, it
  remains true for at least duration $D$. This pattern makes sense
  only if $E_1$ is a predicate on states (that is, on the marking or
  store); since transitions are instantaneous, they have no
  duration.}


\fichepattern
{\code{Absent} $E_1$ \code{before} $E_2$ \code{for duration} $D$}
{No $E_1$ can occur less than $D$ u.t. before the first occurrence of~$E_2$. The pattern holds if there are no occurrence of $E_2$.
}

\fichepattern
{$E_1$ \code{leadsto} first $E_2$ \code{within} $I$ \code{before} $E_3$}
{Before the first occurrence of $E_3$, each occurrence of $E_1$ is
  followed by an occurrence of $E_2$ which occurs both before $E_3$, and in the time
  interval $I$ after $E_1$. The pattern holds if $E_3$ never occurs.}

\fichepattern
{$E_1$ \code{leadsto first} $E_2$ \code{within} $I$ \code{after} $E_3$}
{Same than with the pattern ``$E_1$ \pop{leadsto first} $E_2$ \pop{within} $I$''
   but only considering occurrences of $E_1$ after the first $E_3$.
}

\subsection{Interpretation of Patterns} 
\label{sec:interpr-patt}
We can use different formalisms to define the semantics of
patterns. In this work, we focus on a denotational interpretation,
based on first-order formulas over timed traces (with equality and
trace composition). We illustrate our approach using the pattern $E_1
\pop{leadsto} E_2 \pop{within} I$. 

For the ``denotational'' definition, we say that the pattern $E_1
\pop{leadsto} E_2 \pop{within} I$ is true for a TTS $N$ if and only
if, for every timed-trace $\s$ of $N$, we have:
\[
\forall \s_1, \s_2 \such (\s = \s_1 E_1 \s_2) \Rightarrow \left
  (\begin{array}[c]{@{}l@{}}
    \exists \s_3,\s_4 \such \s_2 = \s_3 E_2 \s_4\\ 
    \quad \wedge \Delta(\s_3)
    \in I \wedge E_2 \notin \s_3
\end{array} \right )
\]
where $\Delta(\s_3)$ is the sum of all the duration in $\s_3$. The
denotational approach is very convenient for a ``tool developer'' (for
instance to prove the soundness of an observer implementing a pattern)
since it is self-contained.

For another example, the denotational definition for the pattern
$\pop{absent} E_2 \pop{after} E_1 \pop{for interval} I$ is given by
the following condition on the traces $\s$ of a system:
\[
\begin{array}[c]{l@{}l}
  \forall \s_1, \s_2, \s_3 \such & (\s = \s_1 E_1 \s_2 E_2 \s_3)\\
  & \   \wedge (E_1 \notin \s_1) \Rightarrow (\Delta(\s_2) \notin I)
\end{array}
\]

On our complete catalogue of patterns~\citep{FRP11}, we provide an
alternative (equivalent) semantics for patterns based on MTL, a timed
extension of linear temporal logic \citetext{see e.g.~\citealp{MITL}
  for a definition of the logic}. For instance, for the \code{leadsto}
pattern, the equivalent MTL formula is $\ltlall\big(E_1 \Rightarrow
((\neg E_2) \until_I E_2)\big)$, which reads like a LTL formula enriched
by a time constraint on the until modality $\until$.

\section{Patterns Verification}
\label{sec:real-time-properties}

We define different types of observers at the TTS level that can be
used for the verification of patterns. It is important to note that we
do not give an automatic method to generate observers. Rather, we
define a set of observers for each patterns and, after selecting the
``most efficient one'', we prove that it is correct (see the
discussion in Sect.~\ref{sec:experimental-results}).  We make use of
the whole expressiveness of the TTS model to build observers:
synchronous or asynchronous rendez-vous (through places and
transitions); shared memory (through data variables); and
priorities. We believe that an automatic method for generating the
observer, while doable, will be detrimental for the performance of our
approach. Moreover, when compared to a ``temporal logic'' approach, we
are in a more favorable situation because we only have to deal with a
finite number of patterns.

\subsection{Observers for the Leadsto Pattern}
\label{sec:observ-leadsto-patt}

We focus on the example of the \code{leadsto} pattern.  We assume that
some events of the system are labeled with $E_1$ and some others with
$E_2$. We give three examples of observers for the
pattern: $E_1$ \code{leadsto} $E_2$ \code{within}
$[0,\mathit{max}[$. The first observer monitors transitions and uses a
single place; the second observer monitors places; the third observer
monitors shared, boolean variables injected into the system (by means
of composition). We define our TTS observers using a classical
graphical notation for Petri Nets, where arcs with a black circle
denote \emph{read arcs}, while arcs with a white circle are
\emph{inhibitor arcs}. (These extra categories of arcs can be defined
in TTS and are supported in our tool chain.)  The use of a \emph{data
  observer} is quite new in the context of TTS systems. The results of
our experiments seem to show that, in practice, this is the best
choice to implement an observer.

\subsubsection{Transition Observer}

\begin{figure}[t]
\begin{center}
      \begin{tikzpicture} [node distance=4.5ex and 7.5ex, label distance=-0.5ex]
    
        \node[transition, label=above:$E_1$] (t1) {}; 
      
        \node[transition, right=of t1, label=below:error, label=above:{$[\mathit{max},\mathit{max}]$}] (t2) {};
      
        \node[transition, right=of t2 , label=above:$E_2$] (t3) {}; 
        \node[place, below=of t2, label=below:obs] (event1) {}
        edge[post] (t2);
        \draw[-stealth] (t1) to[in=125,out=-15] (event1);
        \draw[o-] (t1) to[in=150,out=-50] (event1);
        \draw[o-] (t3) -- (event1);
        \node[transition, below=of t1, label=below:$E_1$] (t4) {}
        edge[readarc] (event1); 
        
        \node[transition, below=of t3, label=below:$E_2$] (t5) {}
        edge[pre] (event1);
        
        \draw[prio] (t2) -- (t5) ;
        
      \end{tikzpicture}
      \caption{Transition Observer: $O_t$}\label{fig:dummy-1}
\end{center}
\end{figure}
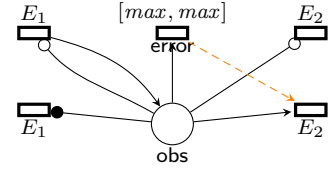

The observer $O_t$, see Fig.~\ref{fig:dummy-1}, uses a place,
\lbl{obs}, to record the time since the last transition $E_1$
occurred. The place \lbl{obs} in $O_t$ is emptied if a transition
labeled $E_2$ is fired, otherwise the transition \lbl{error} is fired
after $\mathit{max}$ unit of time. The priority arc (dashed arrow)
between \lbl{error} and $E_2$ is used to observe the transition
\lbl{error} even in the case where a transition $E_2$ occurs exactly
$\mathit{max}$ u.t. after the place \lbl{obs} was filled.

By definition of the TTS composition operator, the composition of the
observer $O_t$ with the system $N$ duplicates each transitions in $N$
that is labeled $E_1$: one copy can fire if \lbl{obs} is empty---as a
result of the inhibitor arc---while the other can fire only if the
place is full. As a consequence, in the TTS $N \comp O_t$, the
transition \lbl{error} can fire if and only if the place \lbl{obs}
stays full---there has been an instance of $E_1$ but not of
$E_2$---for a duration of $\mathit{max}$. Then, to prove that $N$
satisfies the \code{leadsto} pattern, it is enough to check that the
system $N \comp O_t$ cannot fire the transition \lbl{error}. This can
be done by checking the LTL formula $\ltlall (\neg \code{error})$ on
the system $N \comp O_t$.
%
%

The observer $O_t$ given in Fig.~\ref{fig:dummy-1} is
\emph{deterministic} and will ``react'' to the first occurrence of
$E_2$ that miss a deadline.  It is also possible to define a
non-deterministic observer, such that some occurrences of $E_1$ or
$E_2$ may be disregarded. This approach is safe since model-checking
performs an exhaustive exploration of the states of the system; it
considers all possible scenarios. This non-deterministic behaviour is
quite close to the treatment obtained when compiling an (untimed) LTL
formula ``equivalent'' to the \code{leadsto} pattern, namely $\ltlall
(E_1 \Rightarrow \ltlexist E_2)$, into a Büchi
automaton~\citep{FBAT}. We have implemented the deterministic and
non-deterministic observers and compared them taking in account their
impact on the size of the state graphs that need to be generated and
on the verification time.  Experiments have shown that the
deterministic observer is more efficient, which underlines the benefit
of singling out the best possible observer and looking for specific
optimisation.
    
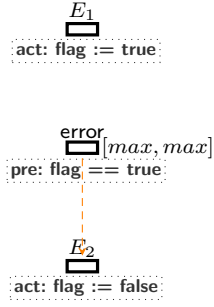
\begin{figure}
\begin{center}
    \begin{tikzpicture} [node distance=7.5ex and 10.5ex, label distance=-0.5ex]
      \node[transition, label=below:{\postact{flag := true}}, label=above:$E_1$] (t1) {};
      \node[transition, below=of t1 , label=above:error, 
      label=below:{\begin{tabular}{c}\preact{flag ==
            true}\\\end{tabular}}, label=right:{$[max,max]$}] (t2) {};
      \node[transition, label=below:{\postact{flag := false}}, below=of t2, label=above:$E_2$] (t3) {};
      \draw[prio] (t2) -- (t3) ;
    \end{tikzpicture}
    \caption{Data Observer: $O_d$}\label{fig:dummy-3}
\end{center}
\end{figure}

\subsubsection{Data Observer}

We define the data observer $O_d$ in Fig.~\ref{fig:dummy-3}. The data
observer has a transition \lbl{error} conditioned by the value of a
boolean variable, \lbl{flag}, that ``takes the role'' of the place
\lbl{obs} in $O_t$ (every boolean variable is considered to be
initially set to false). Indeed, \lbl{flag} is true between an
occurrence of $E_1$ and the following transition $E_2$. Therefore,
like in the previous case, to check if a system $N$ satisfies the
pattern, it is enough to check the reachability of the event
\lbl{error}. Notice that the whole state of the data observer is
encoded in its store, since the underlying net has no place.

\subsubsection{Place Observer} 

\begin{figure}[h]
 \begin{minipage}[b]{0.85\linewidth}\centering
     \begin{tikzpicture} [node distance=4.5ex and 7.5ex, label distance=-0.5ex]
       \node[place, label=below:$S_1$] (event1) {} ;
       \node[transition,  right=of event1, label=right:{$[0,0]$}, label=above:$\tau_1$] (t1) {}
       edge[readarc] (event1);
       \node[place, below=of t1, label=left:obs] (beginobserver) {};
       \node[vtransition, right=of beginobserver , label=below:error, label=above:{$[max,max]$}] (t2) {}
       edge[pre] (beginobserver) ;
       \node[vtransition, below=of beginobserver,  label=right:{$[0,0]$}, label=below:$\tau_2$] (t3) {}
       edge[pre] (beginobserver); 
       \draw[-stealth] (t1) to[in=50,out=-50] (beginobserver);
       \draw[o-] (t1) -- (beginobserver);
       \node[place, left=of t3, label=below:$S_2$] (event2) {};
       \node[vtransition, below=of beginobserver] (t3) {}
       edge[readarc] (event2);
       \draw[prio] (t2) -- (t3) ;
     \end{tikzpicture}
     \caption{Place Observer: $O_p$}\label{fig:dummy-2}
     \end{minipage}
\end{figure}
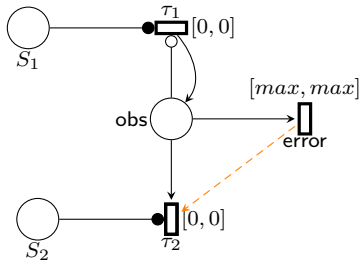

We define the place observer $O_p$ in Fig.~\ref{fig:dummy-2}. In this
section, to simplify the presentation, we assume that the events $E_1$
and $E_2$ are associated to the system entering some given states
$S_1$ and $S_2$. (But we can easily adapt this net to observe events
associated to transitions in the system.) We also rely on a
composition operator that composes TTS through their places instead of
their transitions~\citep{OCTPN} and that is available in our tool
chain.  In $O_p$, we use a transition labeled $\tau_1$ whenever a
token is placed in $S_1$ and a transition $\tau_2$ for observing that
the system is in state $S_2$ (we assume that the labels $\tau_1$ and
$\tau_2$ are fresh---private to the observer---and should not be
composed with the observed systems). The remaining component of $O_p$
is just like the transition observer. We consider both a place and a
transition observer since, depending on the kind of events that are
monitored, one variant may be more efficient than the other.

\subsection{Proving Innocuousness and Soundness of Observers}
\label{sec:prov-innoc-observ}
The goal of this section is to show how to prove that an observer for
a pattern is correct. We demonstrate our approach on the particular
examples of observers for the pattern $E_1 \pop{leadsto} E_2
\pop{within} [0, \mathrm{max}[$, given in the previous section.

We say that an observer $O$ for this pattern is \emph{sound} if it can
``detect'' the traces of a system $N$ that do not hold for the
pattern. More formally, if there is a trace $\s$ of $N$ such that: $\s
= \s_1 E_1 \s_2 E_2 \s_3$ with $\Delta(\s_2) \geq \mathrm{max}$ and
$E_2 \notin \s_2$, then there should be a trace $\s'$ in $N \comp O$
such that $\lbl{error} \in \s'$. (The condition on the trace $\s$
directly follows from the denotational definition of the pattern, see
Sect.~\ref{sec:interpr-patt}) On the opposite, the observer is
\emph{correct} if it can detect that a system satisfies a pattern: if
for all trace $\s'$ of $N \comp O$ we have $\lbl{error} \notin \s'$
then for all trace $\s$ of $N$ the pattern holds.

From our compositionality theorem, see Sect.~\ref{sub:semantics}, we
have that every trace $\s'$ of $N \comp O$ can be defined as the
composition $\s \comp \s_o$ of a trace $\s$ of the system $N$ with a
trace $\s_o$ of the observer $O$. Therefore, to prove that an observer
is correct, it is enough to prove that the pattern does not hold for a
trace $\s_o$ in $O$ iff $\lbl{error} \in s_o$. Indeed, if there is a
trace $\s$ in $N$ that does not hold for the pattern, then we obtain a
trace $\s \comp \s_o$ in $N \comp O$ that does not hold either.

We can use our formal framework to prove the soundness of an observer
(work is currently under way to mechanise these proofs using the Coq
interactive theorem prover). Correctness proofs are more complicated,
since they require to reason on the traces of a system composed with
the observer to figure out the behaviour of the system
alone. Therefore, instead of proving that an observer is correct, we
prove a stronger assumption, that is that observers should be
\emph{innocuous}. A net is said to be \emph{innocuous} if it cannot
interfere with a system placed in parallel. More formally, the TTS $O$
is innocuous if for all TTS $N$ and for all trace $\s$ in $N$ there
exists a trace $\s_o$ in $O$ such that $\s \comp \s_o$ is a trace in
$N \comp O$. Innocuousness means that the observer cannot restrict the
behaviour of another system. This is particularly useful in our case
since, with innocuous observer, any trace $\s$ of the observed
system $N$ is preserved in the composed system $N \comp \obs$: the
observer does not obstruct the behaviour of the system (see
Lemma~\ref{lem/non-intrusive} below).

Instead of proving that observers are non-intrusive in a case by case
basis, we can give a set of sufficient conditions for an observer
$\obs$ to be innocuous. These conditions are met by the three
observers given in Fig.~\ref{fig:dummy-1}--\ref{fig:dummy-2}.

Given a TTS $N$, we say that a transition $t$ of the observer is
\emph{synchronised} when there exists a labeled transition $t'$ of $N$
such that $L(t) = L(t')$ (and the label $L(t)$ is not $\epsilon$).  We
write $\sync$ the set of synchronised transitions of the observer and
$\lsync$ the labels of the synchronised transitions. The transitions
in $\sync$ are the transitions used by the observer to probe the
system. In the examples defined in the previous section, the only
synchronised transitions are the ones labeled $E_1$ and $E_2$ in the
data ($O_d$) and transition ($O_t$) observers.
We define $\imm$ as the set of transitions of the observer whose static
time interval is $[0, 0]$. By construction, no transiton in $\sync$ can also be part of
$\imm$.

\begin{lemma}
\label{lem/non-intrusive}
Assume $\obs$ satisfies the following three conditions:
\begin{itemize}
\item all synchronised transitions have a trivial static time interval
  and no priority (that is, for every $t$ in $\sync$, $I_s^t = [0 ;
  +\infty[$ and $t$ has no priority over another transition in $O$);

\item from any state of the observer, and for every label $l \in
  \lsync$, there is at least one transition $t$ in $\obs$ with label
  $l$ that can fire immediately;

\item from any state of the observer, there is no infinite sequence of
transitions in $\imm$.

\end{itemize}
then, for all timed trace $\s$ in $N$ there exists a timed trace $\s
\comp \s_o$ in $N \comp \obs$ such that $\s_o$ is a trace of $\obs$.
\end{lemma}

The proof of Lemma~\ref{lem/non-intrusive} can be found
in~\citet{VRTS11}. A few comments on these conditions. The first
condition is necessary for defining the composition of two TTS (see
Sect.~\ref{sub:semantics}). The second condition ensures that the
observer cannot delay the firing of a synchronised transition ``for a
non-zero time''.  Assume $s$ is a state of the observer $\obs$ and
$\s$ a finite trace of $\obs$ starting from state $s$. We define
$\obs(s, \s)$ to be the (necessarily unique) state reached by the
observer after trace $\s$ has been executed. From the second
condition, in every reachable state $s$ of $\obs$, and for every label
$l$ in $L(\sync)$, there exists a (possibly empty) finite trace $\s$
not containing transitions in $\sync$ such that the duration of $\s$
is 0 and there exists $t \in \sync$ with $L(t) = l$, which is fireable
in state $\obs(s, \s)$. Note also that the observer cannot involve
other synchronised transitions while reaching a state where $l$ is
firable, since this would abusively constrain the behaviour of the main
system $N$, not to mention deadlock issues. This condition is true for
the observer $O_t$ in Fig.~\ref{fig:dummy-1} since, at any time,
exactly one of the two transitions labeled $E_1$ (resp. $E_2$) can
fire.


\section{Experimental Results}
\label{sec:experimental-results}
Our verification framework has been integrated into a prototype
extension of \emph{frac}, the Fiacre compiler for the TINA toolbox.
This extension supports the addition of real time patterns and
automatically compose a system with the necessary observers. (Software
and examples are available at \url{http://homepages.laas.fr/~nabid}.)
In case the system does not meet its specification, we obtain a
counter-example that can be converted into a timed sequence of events
exhibiting a problematic scenario. This sequence can be played back
using two programs provided in the TINA tool set, \emph{nd} and
\emph{play}. The first program is a graphical animator for Time Petri
Net, while the latter is an interactive (text-based) animator for the
full TTS model.

We define the \emph{empirical complexity} of an observer as its impact
on the augmentation of the state space size of the observed
system. For a system $S$, we define $\mathit{size}(S)$ as the size (in
bytes) of the \emph{State Class Graph} (SCG)~\citep{tina} of $S$
generated by our verification tools. In TINA, we use SCG as an
abstraction of the state space of a TTS.  State class graphs exhibit
good properties: an SCG preserves the set of discrete traces---and
therefore preserves the validation of LTL properties---and the SCG of
$S$ is finite if the Petri Net associated with $S$ is bounded and if
the set of values generated from $S$ is finite. We cannot use the
``plain'' labeled transition system associated to $S$ to define the
size of $S$; indeed, this transition graph maybe infinite since we
work with a dense time model and we have to take into account the
passing of time.

The size of $S$ is a good indicator of the memory footprint and the
computation time needed for model-checking the system $S$: the time
and space complexity of the model-checking problem is proportional to
$\mathit{size}(S)$. Building on this definition, we say that the
complexity of an observer $O$ applied to the system $S$, denoted
$C_O(S)$, is the quotient between the size of $(S \comp O)$ and the
size of $S$.

\begin{figure*}[htb]
\centering
\includegraphics[height=6cm]{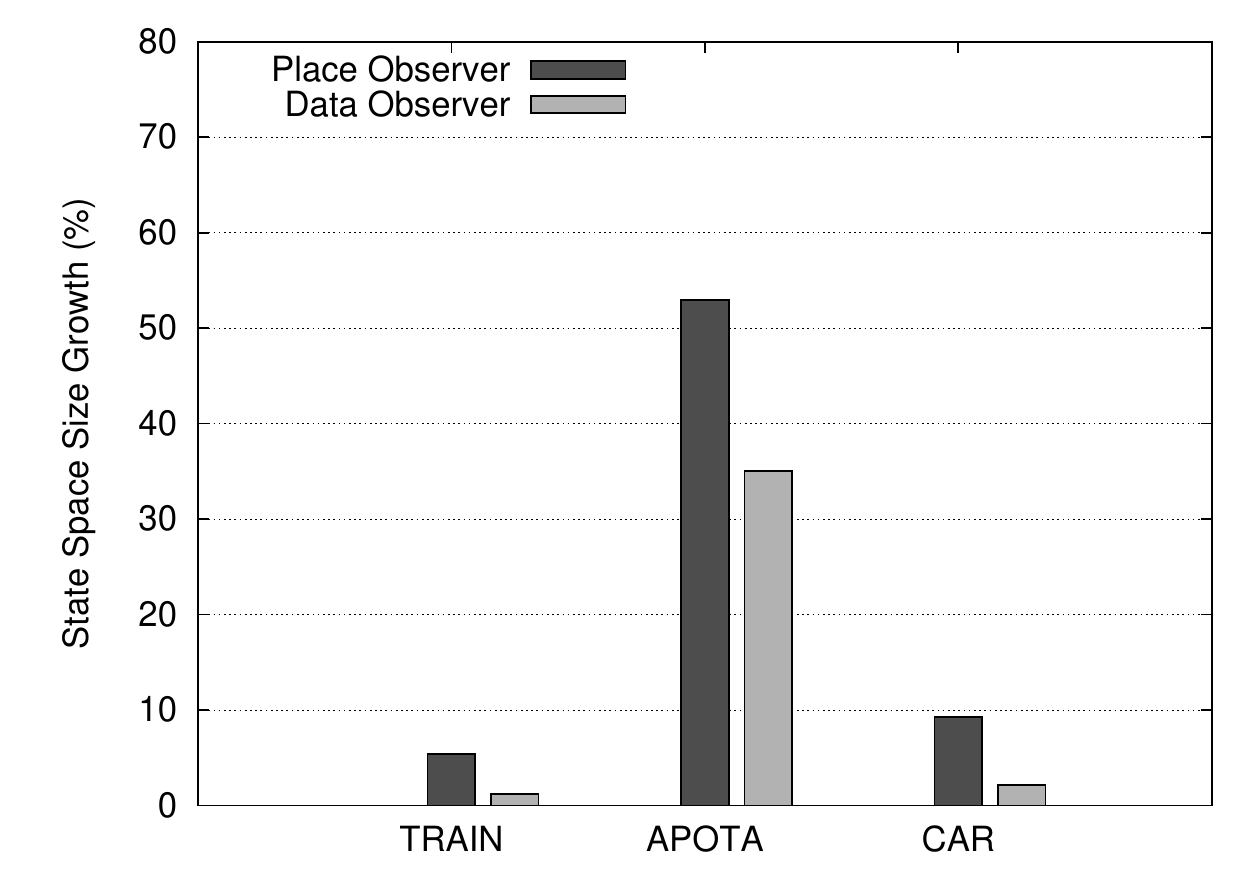} \hfill
\includegraphics[height=6cm]{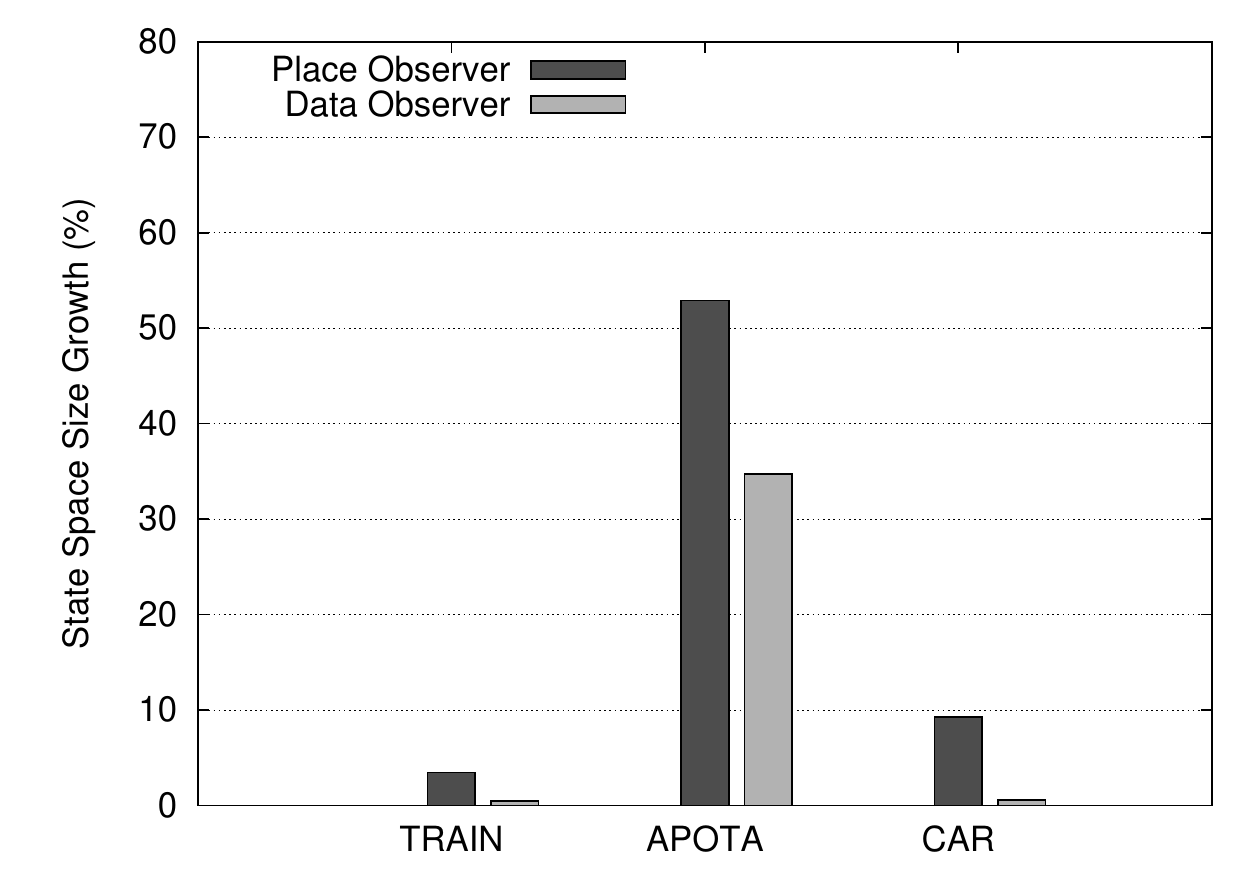}
\caption{Compared complexity of the data and place observers (in
  percentage of system size growth)---average time for invalid
  properties (right) and valid properties (left).}
\label{fig:example}
\end{figure*}


We resort to an empirical measure for the complexity since we cannot
give an analytical definition of $C_O$ outside of the simplest
cases. However, we can give some simple bounds on the function
$C_O$. First of all, since our observers should be non-intrusive, we can show that the SCG of $S$ is
a sub graph of the SCG of $S \comp O$, and therefore $C_O(S) \geq
1$. Also, in the case of the \code{leadsto} pattern, the transitions
and places-based observers add exactly one place to the net associated
to $S$. In this case, we can show that the complexity of these two
observers is always less than~$2$; we can at most double the size of
the system. We can prove a similar upper bound for the \code{leadsto}
observer based on data. While the three observers have the same
(theoretical) worst-case complexity, our experiments have shown that
one approach was superior to the others. We are not aware of previous
work on using experimental criteria to select the best observer for a
real-time property. In the context of ``untimed properties'', this
approach may be compared to the problem of optimising the generation
of Büchi Automata from LTL formulas, see e.g.~\citet{FBAT}.

We have used our prototype compiler to experiment with different
implementations for the observers. The goal is to find the most
efficient observer ``in practice'', that is the observer with the
lowest complexity. To this end, we have compared the complexity of
different implementations on a fixed set of representative examples
and for a specific set of properties (we consider both valid and
invalid properties). The results for the \code{leadsto} pattern are
displayed in Fig.~\ref{fig:example}. For the experiments used in this
paper, we use three examples of systems selected because they exhibit
very different features (size of the state space, amount of
concurrency and symmetry in the system, \dots):
\begin{itemize}
\item TRAIN is a model of a train gate controller. The example models
  a system responsible for controlling the barriers protecting a
  railroad crossing gate. When a train approaches, the barrier must be
  lowered and then raised after the train's departure. The valid
  property, for the TRAIN example, states that the delay between
  raising and lowering a barrier does not exceed 100 unit of time. For
  the invalid property, we use the same requirement, but shortening
  the delay to 75.
\item APOTA is an industrial use case that models the dynamic
  architecture for a network protocol in charge of data communications
  between an air plane and ground stations~\citep{FVAMFT}. This
  example has been obtained using an translation from AADL to
  Fiacre. In this case, timing constraints arise from timeouts between
  requests and periods of the tasks involved in the protocol
  implementation. The property, in this case, is related to the
  worst-case execution time for the main application task.
\item CAR is a system modelling an automated rail car system taken
  from~\citet{tap}. The system is composed of four terminals connected
  by rail tracks in a cyclic network. Several rail cars, operated from
  a central control center, are available to transport passengers
  between terminals. When a car approaches its destination, it sends a
  request to the terminal to signal its arrival. Passengers in the
  terminal can then book a travel in the car. The valid property, for
  the CAR example, states that a passenger arriving in a terminal,
  must have a car ready to transport him within 15\, unit of time. For
  the invalid property, we use the same requirement, but shortening
  the delay to 2 unit of time.
\end{itemize}

In Fig.~\ref{fig:example}, we compare the growth in the state space
size---that is the value of $C_o(S)$---for the place and data
observers defined in Sect.~\ref{sec:observ-leadsto-patt} and our three
running examples. We do not consider the transition observer in these
results since the events used in the requirements are all related to a
system entering a state (and therefore our benchmark favor the place
observer over the transition observer). We use one chart to display
the result for patterns that are invalid and another for valid
patterns.

In Fig.~\ref{fig:example2} (page~\pageref{fig:example2}), we give
results on the total verification time for the APOTA example. 
The
value displayed in the table refer to the time spent generating the
complete state space of the system and verifying the property. The row
SYSTEM gives the time needed for exploring the complete state space of
the system (without adding any observer) while ``VALID'' and
``INVALID'' refer to the state space of the system synchronised with
data observer and state observer in the case of valid and invalid
property respectively.

In our experiments, we have consistently observed that the observer
based on data is the best choice; it is the observer giving the
minimal execution time in almost all the cases and that seldom gives
the worst result. 
We can explain the efficiency of the data observer by the fact that it
adds less transitions than the state observer; which means that it
adds less intermediary states to the state space of $(N \comp O)$.

\begin{figure}[htb]
  \centering
\begin{tabular}[c]{|p{0.15\textwidth}|p{0.1\textwidth}|p{0.1\textwidth}|}
\hline
\centering
Example & State observer & Data observer \\
\hline
SYSTEM & \hfill 2.861 & \hfill 2.861\\
\hline
VALID & \hfill 11.662 & \hfill 10.652\\
\hline
INVALID & \hfill 11.611 & \hfill 10.179\\
\hline
 \end{tabular}
 \caption{Total verification time for APOTA (in seconds)}
 \label{fig:example2}
 \end{figure}


\section{Related Work}
Two broad approaches coexist for the definition and verification of
real-time properties: (1) real-time extensions of temporal logic~\citep{H98}; and
(2) observer-based approaches, such as the Context Description Languages
(CDL) of Dhaussy et al.~\citep{ECOVFM} or approaches based on timed
automata~\citep{MITL,MCRTTA,TPRTTA}.

Obviously, the logic-based approach provides most of the theoretically
well-founded body of works, such as complexity results for different
fragments of real-time temporal logics~\citep{H98}: Temporal logic with
clock constraints (TPTL); Metric Temporal Logic---with or without
interval constrained operators---; Event Clock Logic; etc. The
algebraic nature of logic-based approaches make them expressive and
enable an accurate formal semantics. However, it may be impossible to
express all the necessary requirements inside the same logic fragment
if we ask for an efficient model-checking algorithm (with polynomial
time complexity). For example, Uppaal~\citep{behrmann04atutorial} chose
a restricted fragment of TCTL with clock variables, while Kronos
provide a more expressive framework, but at the cost of a much higher
complexity. As a consequence, selecting this approach requires to
develop model-checkers for each interesting fragment of these
logics---and a way to choose the right tool for every
requirement---which may be impractical.

Pattern-based approaches propose a user-friendly syntax that
facilitates their adoption by non-experts. However, in the real-time
case, most of these approaches lack in theory or use inappropriate
definitions. One of our goal is to reverse this situation. In the
seminal work of~\citet{ppsfsv1999}, patterns are defined by
translation to formal frameworks, such as LTL and CTL. There is no
need to provide a verification approach, in this case, since efficient
model-checkers are available for these logics. This work on patterns
has been extended to the real-time case. For example,~\citet{RSP} extends the patterns language with time constraints
and give a mapping from timed pattern to TCTL and MTL, but they do not
study the decidability of the verification method (the
implementability of their approach). Another related work
is~\citep{PTPS}, where the authors define observers based on Timed
Automata for each pattern. However, they do not provide a formal
framework for proving the correctness or the innocuousness of their
observers and they have not integrated their approach inside a
model-checking tool chain.


Concerning observer-based approaches, \citet{TPRTTA,MCRTTA} use test automata to check
properties of reactive systems. The goal is to identify properties on
timed automata for which model checking can be reduced to reachability
checking. In this framework, verification is limited to safety and
bounded liveness properties. In the context of Time Petri Net, a
similar approach has been experimented by~\citet{TCVMBTPN}, but they propose a less general model for
observers and consider only two verification techniques over four
kinds of time constraints. \citet{APTABI} propose a method 
to verify the correctness of their approach formally. However,
they do not prove formally all their invariants (patterns in
our case).

\section{Contributions and Perspectives}
\label{sec:contr-persp}

In contrast to these related works, we make the following
contributions. We reduce the problem of checking real-time properties
to the problem of checking LTL properties on the composition of the
system with an observer. We define also a real-time patterns language
based on the work of~\citet{ppsfsv1999} and inspired from real-case
studies.  To choose the best way to verify a pattern, we defined, for
each pattern, a set of non-intrusive observers.  We are based on a
formal framework to verify the correctness of an observer, whether it
can interfere with the behaviour of the system under observation.

Our approach has been integrated into a complete verification
tool chain for the Fiacre modelling language and can therefore be used
in conjunction with Topcased~\citep{ttptosfcasd2006}.  We give several experimental results
based on the use of this tool chain in
Sect.~\ref{sec:experimental-results}. The fact that we implemented our
approach has influenced our definition of the observers. Indeed,
another contribution of our work is the use of a pragmatic approach
for comparing the effectiveness of different observers for the same
property. Our experimental results seem to show that data observers
look promising.

We are following several directions for future work. A first goal is
to define a new low-level language for observers---adapted from the
TTS model---equipped with more powerful optimisation techniques and
with easier soundness proofs. On the theoretical side, we are
currently looking into the use of mechanised theorem proving
techniques to support the validation of observers. On the experimental
side, we need to define an improved method to select the best
observer. For instance, we would like to provide a tool for the
``syntax-directed selection'' of observers that would choose (and even
adapt) the right observers based on a structural analysis of the
target system.

\end{document}